\documentclass[aps,preprintnumbers,prl,twocolumn,superscriptaddress,nofootinbib]{revtex4}

\usepackage{epsfig,latexsym,cancel,amssymb,amsmath,adjustbox}
\usepackage{graphicx}
\usepackage{epstopdf}
\usepackage{hyperref, graphicx, slashed,  amsmath}

\usepackage[dvipsnames]{xcolor}

\allowdisplaybreaks

\newcommand{\inab}{\,{\rm ab}^{-1}}
\newcommand{\infb}{\,{\rm fb}^{-1}}

\newcommand{\eeww}{e^+e^- \to WW}

\newcommand{\A}{\mathcal{A}}
\newcommand{\M}{\mathcal{M}}

\newcommand{\bpm}{\begin{pmatrix}}
\newcommand{\epm}{\end{pmatrix}}

\newcommand{\la}{\langle}
\newcommand{\ra}{\rangle}

\newcommand{\eeaa}{e^+e^- \to \gamma \gamma}
\newcommand{\mumuaa}{\mu^+\mu^- \to \gamma \gamma}

\newcommand{\eemmaa}{e^+e^- (\mu^+\mu^-) \to \gamma \gamma}

\newcommand{\Affaa}{\mathcal{A}(f^+f^- \gamma^+ \gamma^-)}

\renewcommand{\sec}[1]{{\textbf{\textit{#1}}}.---}
\renewcommand{\sec}[1]{{{\textit{#1}}}.---}

\begin{document}

\preprint{MITP/20-063}

\title{Unambiguously Testing Positivity at Lepton Colliders}

\author{Jiayin Gu}
\email{jiayin\_gu@fudan.edu.cn}
\affiliation{Department of Physics and Center for Field Theory and Particle Physics, Fudan University, Shanghai 200438, China}
\affiliation{Key Laboratory of Nuclear Physics and Ion-beam Application (MOE), Fudan University, Shanghai 200433, China}
\affiliation{PRISMA$^+$ Cluster of Excellence, Institut f\"ur Physik,
Johannes Gutenberg-Universit\"at, Staudingerweg 7, 55128 Mainz, Germany}
\author{Lian-Tao Wang}
\email{liantaow@uchicago.edu}
\affiliation{Department of Physics and Enrico Fermi Institute, University of Chicago, Chicago, IL 60637, USA}
\affiliation{Kavli Institute for Cosmological Physics, University of Chicago, Chicago, IL 60637, USA}
\author{Cen Zhang}
\email{cenzhang@ihep.ac.cn}\thanks{corresponding author}
\affiliation{
Institute for High Energy Physics, and School of Physical Sciences, University
of Chinese Academy of Sciences, Beijing 100049, China
}
\affiliation{Center for High Energy Physics, Peking University, Beijing 100871, China}

\begin{abstract}
The diphoton channel at lepton colliders, $\eemmaa$, has a remarkable feature that the leading new physics contribution comes only from dimension-eight operators.  
This contribution is subject to a set of positivity bounds,  
derived from the fundamental principles of Quantum Field Theory, such as 
unitarity, locality, analyticity and Lorentz invariance.
These positivity bounds are thus applicable to the most direct observable --- the diphoton cross section. This unique feature provides a clear, robust, and unambiguous test of these principles. 
We estimate the capability of various future lepton colliders in probing the dimension-eight operators and testing the positivity bounds in this channel.  
We show that positivity bounds can lift certain flat directions among the effective operators and significantly change the perspectives of a global analysis.   
We also discuss the positivity bounds of the $Z\gamma/ZZ$ processes which are related to the $\gamma\gamma$ ones, but are more complicated due to the massive $Z$ boson.  \\

\begin{center}
{\it  In memory of Cen Zhang}
\end{center}

\end{abstract}

\maketitle

\sec{Introduction} 
Positivity bounds on coefficients for operators in the Standard Model Effective
Field Theory (SMEFT) arise from the assumption that
their UV completion obeys the fundamental principles of quantum field theory (QFT),
such as unitarity, locality, analyticity and Lorentz invariance.  Testing these bounds at colliders
is difficult, as they generally only apply to effects of dimension-eight
(dim-8) or higher operators~\cite{Adams:2006sv,  Bellazzini:2016xrt,
deRham:2017avq, deRham:2017zjm, Chandrasekaran:2018qmx, deRham:2018qqo,
Bellazzini:2018paj, Bi:2019phv, Remmen:2019cyz, Remmen:2020vts, Zhang:2020jyn,
Fuks:2020ujk, Yamashita:2020gtt}.  Their contribution to a given process is
typically subleading compared with those from dimension-six (dim-6) operators,
making their measurements experimentally challenging even with differential
observables~\cite{Alioli:2020kez, Fuks:2020ujk}.  For dim-6 operators, such
bounds do not exist without explicit assumptions on the UV
model~\cite{Low:2009di, Falkowski:2012vh, Bellazzini:2014waa, Gu:2020thj,
Remmen:2020uze}.

In this letter, we identify a specific process, $\eeaa$ (or $\mumuaa$), in
which the dim-8 operators provide the leading new physics contribution, and a
test of the positivity bounds can be unambiguously carried out.  The
measurements of this simple process at lepton colliders thus have profound
implications.  A confirmed violation of the positivity bounds would be more
revolutionary than any particle discovery, as it
would indicate a breakdown of at least one of the foundations of QFT.

\sec{The diphoton channel}
The leading new physics contributions to $\eeaa$ appear at dim-8.  
This can be easily deduced in the massless tree-level limit as follows, and 
we postpone a more detailed discussion of the dim-6 effects  
to the next section.   Neglecting the electron mass, the tree-level SM amplitude
takes only the $\Affaa$ helicity configuration (the superscripts denote the helicity), where $f=e_{L,R}$ is the left- or right-handed electron.    The lowest order new physics contribution to the same helicity amplitude (required to generate an interference term with the SM) is a contact interaction that 
has mass dimension four, which is generated by dim-8 operators.   
Denoting with $e$ the electric coupling, $v$ the Higgs vacuum expectation value (vev),
the amplitude of the diphoton process can be written as
%
%
%
\begin{align}
&~ \Affaa_{\rm SM + d8} \nonumber\\ =&~   2e^2 \frac{\la 24 \ra^2}{ \la 13 \ra  \la 2 3 \ra } + \frac{a}{v^4} [13][23] \la 24 \ra^2       \nonumber\\ 
=&~  2e^2 \frac{\la 24 \ra^2}{ \la 13 \ra  \la 2 3 \ra } \left( 1 +  \frac{a}{2 e^2 v^4} tu \right)    \,, 
\label{eq:Aeeaa}
\end{align}
where the effective parameter $a$ (denoted as $a_{L,R}$ later for $f=e_{L,R}$) depends on the dim-8 coefficients, and $s,t,u$ are the Mandelstam variables.   
Here we only highlight the key features of the helicity amplitude formalism used in Eq.~(\ref{eq:Aeeaa}) and refer the readers to recent reviews~\cite{Elvang:2013cua, Dixon:2013uaa, Cheung:2017pzi} for more details. 
The two-component spinor $|p]$ ($|p\ra$) has mass dimension $1/2$ and helicity $+1/2$ ($-1/2$).  The total helicities of the amplitude need to be consistent with the ones of the external particles (labelled in numerical order). This uniquely fixes the form of the dim-8 contact term, which has an overall mass dimension of four, while $[13]^2\la14\ra \la24\ra = -  [13][23] \la 24 \ra^2$ is not independent.  A contact term with a lower mass dimension could not be written down for the same helicity amplitude.   
We note here that, by definition, a positive $a$ indicates a constructive interference between the SM and the dim-8 amplitudes.  

Positivity bounds can be derived from a
twice-subtracted dispersion relation, assuming that
the UV completion obeys the fundamental principles of QFT~\cite{Adams:2006sv}.
The dispersion relation connects the second $s$ derivative of an elastic amplitude to an integration of its discontinuity,
which is positive definite.
Rotating the diphoton amplitude to the elastic process $e \gamma \to e
\gamma$, and taking the forward limit, we have
\begin{align}
\A(f^+\gamma^+ f^- \gamma^-)_{\rm SM + d8} =&~   2e^2 \frac{\la 34 \ra^2}{ \la 12 \ra  \la 3 2 \ra } \left( 1 +  \frac{a}{2 e^2 v^4} su \right)       \nonumber\\ 
\underset{t\to 0}{=}\!&~   \M_{\rm SM}  \left( 1 -  \frac{a}{2 e^2 v^4} s^2 \right)     \,, 
\label{eq:Aeaea}
\end{align}
where $\M_{\rm SM} \equiv  2e^2 \frac{\la 34 \ra^2}{ \la 12 \ra  \la 3 2 \ra } |_{t \to 0}$ is the SM amplitude in the forward limit.  
An important feature of the $e \gamma \to e \gamma$ process is that, in the forward limit where the positivity bound is derived, the SM amplitude is a nonzero finite constant, and one could explicitly show that (see Appendix~\ref{sec:app}) $\M_{\rm SM} = -2e^2$.      
This is in contrast with the examples in Ref.~\cite{Adams:2006sv}, where the dim-4 Lagrangian is a free theory and the interference term does not exist.  In other cases (such as the scattering of two fermions), the SM elastic amplitude may have a $t$-channel pole from the exchange of a massless particle, and the forward limit is not well defined.  In such cases, additional treatments are needed to obtain meaningful positivity bounds, for instance by systematically subtracting all calculable SM contributions to the amplitude before taking the forward limit.  It is also possible for the SM amplitude to have $s$-channel poles that may contaminate the positivity bounds of dim-8 coefficients, which again need to be systematically subtracted.   These cases introduce additional steps and subtleties in understanding the implications of positivity bounds on observables.  
The fact that the SM $e \gamma \to e \gamma$ forward amplitude is a finite constant means that the positivity bound also uniquely fixes the relative sign of the SM and dim-8 contributions, as suggested by \autoref{eq:Aeaea}.  
Due to crossing, a positive $a$ here corresponds to a destructive interference between the SM and dim-8 terms.  
Since $\M_{\rm SM} = -2e^2 <0$, the positivity bound, 
\begin{equation}
\frac{d^2}{ds^2} \A(f^+\gamma^+ f^- \gamma^-)|_{t\to 0} \ge 0 \,,
\end{equation}
 implies $a\ge 0$.  The interference between SM and dim-8 contributions is thus bounded to be destructive in $e \gamma \to e
\gamma$, and {\bf constructive in $\boldsymbol{\eeaa}$}.

One could work in the {\it amplitude basis}~\cite{Ma:2019gtx, Durieux:2019siw}
and directly connect Eq.~(\ref{eq:Aeeaa}) with the massless amplitudes of the $W$
and $B$ fields in the unbroken electroweak phase. Alternatively,
using the basis of Ref.~\cite{Murphy:2020rsh} (see also Ref.~\cite{Li:2020gnx}),
$a_L$ and $a_R$ are given by 
\begin{align}
	a_L =& -2\frac{v^4}{\Lambda^4} \left( c_W^2 c_{l^2B^2D} - 2 s_Wc_W
	c^{(2)}_{l^2WBD} + s_W^2 c^{(1)}_{l^2W^2D} \right) ,
	\nonumber\\
a_R =&-2 \frac{v^4}{\Lambda^4} \left( c_W^2 c_{e^2B^2D}  +  s_W^2 c_{e^2W^2D} \! \right) \,, \label{eq:aLRop}
\end{align}
where $s_W\equiv \sin\theta_W$, $c_W\equiv \cos\theta_W$, $c_i$'s are the
coefficients of the 
five dim-8 operators, $Q_i$, as defined in Ref.~\cite{Murphy:2020rsh} (see Appendix~\ref{sec:app}), and $\Lambda$ denotes the scale of the potential new physics.  
They are the only relevant operators in the full dim-8 basis, 
not only for diphoton but also for the CP-even $\A(\bar{e}_L e_L V_1^+
V_2^-)_{\rm d8}$ and $\A(e_R \bar{e}_R V_1^+ V_2^-)_{\rm d8}$ amplitudes in the massless limit, where
$V_{1,2}=Z,\gamma$.

\sec{Dim-6 contributions}
A dim-6 tree-level contribution to the diphoton process can be generated only by a
dipole operator, and has a different fermion helicity configuration than the SM one, $\Affaa$. The dim-6
interference term therefore does not exist~\cite{Azatov:2016sqh}. 
At the one-loop level, several dim-6 contributions arise, but they are all
strongly constrained by other measurements, and can be safely ignored with 
a loop factor suppression.
For instance,
operator $\mathcal{O}_{3W} = \frac{1}{3!} g \epsilon_{abc} W^{a\,\nu}_\mu
W^b_{\nu \rho} W^{c\,\rho\mu} $ contributes to $\Affaa$ at one loop, but it can
be very-well probed by the $\eeww$ process.  A rough
estimation with the projections from Ref.~\cite{deBlas:2019wgy} 
($\lesssim 10^{-4}$ in terms of the anomalous triple-gauge couplings) 
suggests that
its impact on the diphoton cross section is at most around $\delta
\sigma_{\gamma\gamma}/\sigma_{\gamma \gamma } \sim 10^{-7}$, much smaller than
the expected precision at a realistic lepton collider (see Appendix~\ref{sec:app}).  
%
Similarly, the modifications in the $Ze^+e^-$ ($Z\mu^+\mu^-$) couplings are already stringently constrained at  the $10^{-4}$ ($10^{-3}$) level even with current measurements~\cite{deBlas:2019wgy}, and their loop contributions to the diphoton process can be safely neglected.  The one loop contributions involving the Higgs boson are also irrelevant since they are suppressed by the square of electron (muon) Yukawa coupling. 
%
%
While the four-fermion operators
involving two electrons and two top-quark fields are poorly constrained,
their contribution to $\Affaa$ is forbidden by the angular momentum
selection rules, since they cannot produce the $J=2$ state of two
photons~\cite{Jiang:2020sdh}.  
The interference between the one-loop SM amplitude and the tree-level dipole contribution is also absent with massless electrons.  
Contributions with two insertions of dim-6
operators are formally indistinguishable from dim-8 operators.  At the tree
level, the only such contribution which is not equivalent to a contact dim-8
operator insertion comes from two insertions of electron dipole
couplings~\footnote{One could always choose a basis, such as the Warsaw
	basis~\cite{Grzadkowski:2010es}, in which the only dim-6-squared
contribution comes from the dipole operators.}. 
They are strongly constrained by the $g_e -2$ and the electric dipole moment
measurements~\cite{Andreev:2018ayy, Bennett:2008dy}. 
A rough estimation suggests that their impact on the diphoton cross section is at most $\frac{\delta\sigma_{\gamma\gamma}}{\sigma_{\gamma \gamma }}   \sim (\frac{E}{10^7{\rm TeV}})^2$ where $E$ is the center-of-mass energy, and can be safely ignored.  
Finally, we note that dim-8 operators involving Higgs fields do not contribute to
$\Affaa$ either, 
as the insertion of a Higgs vev makes the amplitude effectively at a lower mass
dimension, where a contact term for $\Affaa$ does not exist.

Naively, one expects that the dim-6 contributions from new physics will be
first observed in some other processes. What then is the motivation to look
for dim-8 deviations in $\eeaa$? 
First, testing positivity at dim-8 provides more fundamental
information about the nature of new physics, namely whether it is consistent
with the QFT framework, which one cannot tell from a SMEFT analysis truncated
at dim-6. 
An observation of dim-6 deviation elsewhere would only strengthen the
motivation to test dim-8 deviations in $\eeaa$.  Second, dim-6 effects from
different UV states could be suppressed due to dynamics \cite{Agashe:2016rle},
certain symmetries~\cite{Gu:2020thj}, or accidental cancellation.  In contrast,
constraining the positively bounded dim-8 effects could lead to
unambiguous exclusion limits on all possible UV particles, as 
each of them contributes positively, assuming the QFT framework is valid
\cite{Fuks:2020ujk}.


\sec{Positivity bounds on cross sections}
The positivity bounds, $a_L\ge0$ and $a_R\ge0$, restrict the interference between SM and dim-8 contributions to be constructive in $\eeaa$. As such, they can be directly related to the cross section.  
Since the helicities of the two photons cannot be measured in practice,  we work with the folded
distribution of the production polar angle $\theta$,
%
\begin{align}
&~\frac{d\sigma(\eeaa)}{d |\!\cos\theta|} \nonumber\\ =&~~~  \frac{(1-P_{e^-})(1+P_{e^+})}{4} \frac{e^4}{4 \pi s} \left(\frac{1+c^2_\theta}{1-c^2_\theta} + a_L \frac{s^2(1+c^2_\theta)}{4 e^2 v^4}   \right) \nonumber\\
& + \frac{(1+P_{e^-})(1-P_{e^+})}{4} \frac{e^4}{4 \pi s} \left(\frac{1+c^2_\theta}{1-c^2_\theta} + a_R \frac{s^2(1+c^2_\theta)}{4 e^2 v^4}  \right) \,,   \label{eq:dsigma1}
\end{align}
where $s$ is the square of the center-of-mass energy, $P_{e^-}$ ($P_{e^+}$) is
the polarization of the electron (positron) beam, and $c_\theta \equiv |\!\cos
\theta|$.  
The $a_L$, $a_R$ terms come from the interference between SM and dim-8 operators,
while the dim-8-squared contributions can be safely neglected due to the high measurement precision of this channel at lepton colliders.
It is now clear that the positivity bounds 
$a_L \ge 0,\,a_R \ge 0$ have a simple consequence, namely
\begin{equation}
\frac{d \sigma}{d|\!\cos\theta|} (\eeaa) \ge  \frac{d  \sigma_{\rm SM} }{d|\!\cos\theta|} (\eeaa) \,, \label{eq:posxaa}
\end{equation}
for any beam polarizations and any $|\!\cos\theta|$.
We see that the $\eeaa$ channel is special in that the positivity of Wilson
coefficients can be directly translated into positivity in realistic
observables, without being contaminated by any other non-positive operators.
The diphoton process thus provides a simple, clear, and unambiguous test of the
fundamental principles of QFT.  

It is also interesting to note that the measurements at LEP2 display an overall
signal strength of $\eeaa$ to be about 1.5 standard deviations below the SM
expectation \cite{Schael:2013ita}.  While statistically insignificant, 
this deviation
exhibits a small tension with the positivity bound.  

\sec{Collider reach} 
To estimate the reach at future lepton
colliders, we perform a simple binned analysis in the range $|\!\cos\theta| \subset[0,0.95]$, with a bin width of $0.05$, and consider only statistical uncertainties. 
A differential analysis in $|\!\cos\theta|$ helps discriminate the dim-8 contribution from the SM one, as the latter dominates the forward region due to the $t/u$-channel electron exchange.
We expect the largest background to be the Bhabha scattering ($e^+e^- \to
e^+e^-$). 
Assuming a sufficiently small rate ($\ll 1\%$) for an electron to be
misidentified as a photon, 
this background is more than two orders of magnitude smaller than the signal (see Appendix~\ref{sec:app}).  
The cut on
the minimal production polar angle ($|\!\cos\theta|<0.95$) is also very effective
in removing the beamstrahlung and ISR effects.  
%
%
We note that the reach on $\Lambda$ is only mildly sensitive to the measurement uncertainties ($\sim \Delta^{-1/4}$) due to the $1/\Lambda^4$ dependence of the dim-8 contribution, so our analysis gives a reasonable projection as long as the systematic uncertainties are not overwhelmingly large. 
As a validation, 
we apply it to the LEP2 run scenarios and find a very good agreement with the
result of Ref.~\cite{Schael:2013ita} (with a $\lesssim 10\%$ difference in the
reach on $\Lambda$).  

\begin{figure}[h!]
\centering
\includegraphics[width=0.3\textwidth]{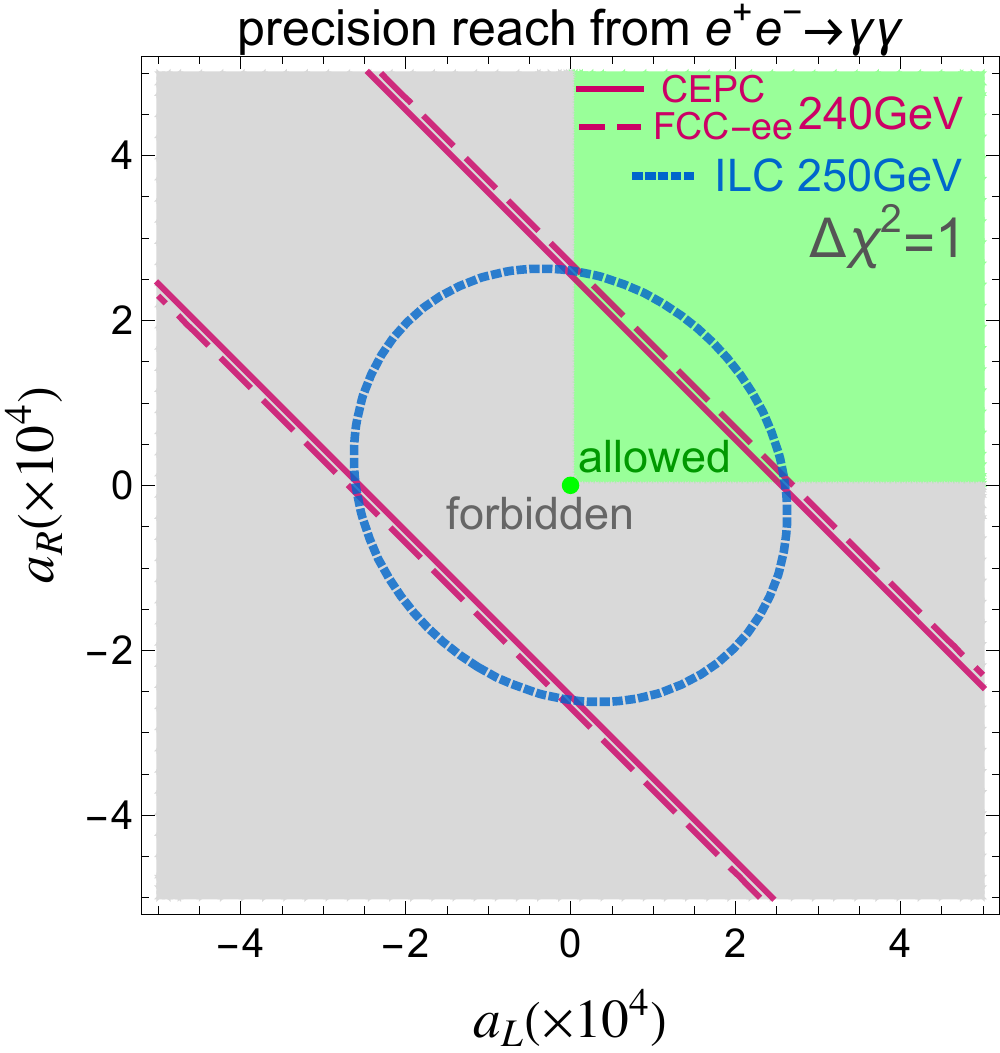}
\caption{$\Delta \chi^2=1$ contours for CEPC/FCC-ee 240\,GeV and ILC 250\,GeV.
	The green shaded region is allowed by the positivity bounds. 
	}
\label{fig:chisaa1}
\end{figure}

To illustrate the interplay between the measurements and the positivity bounds,
we show the $\Delta \chi^2=1$ contours in \autoref{fig:chisaa1} for collider
scenarios 
CEPC/FCC-ee 240\,GeV \cite{CEPCStudyGroup:2018ghi,Abada:2019lih,Abada:2019zxq}
and ILC 250\,GeV \cite{Bambade:2019fyw}. 
According to Eq.~(\ref{eq:dsigma1}), if the beams
are unpolarized ($P_{e^-} = P_{e^+} =0$), only the combination $a_L+a_R$ is
probed, leaving a flat direction along $a_L=-a_R$ as shown by the diagonal band (indicating $\Delta \chi^2\leq 1$) for CEPC/FCC-ee. It can be lifted by having
multiple runs with different beam polarization, as for example at the ILC.
Clearly, beam polarization is 
desirable, because it allows for testing the signs of $a_L$ and $a_R$ (or
the two polarized cross sections) individually. 

On a different ground, assuming the UV completion is consistent
with the QFT principles which imply $a_L,\,a_R \geq 0$, 
%
$a_L$ and $a_R$ can be simultaneously constrained even without beam polarization, 
as illustrated in \autoref{fig:chisaa1}.
%
This is a general feature that also applies to many other processes, such as the
fermion scattering~\cite{Fuks:2020ujk} or the Higgs production~\cite{Hays:2018zze}.  
Positivity thus provides important information for future global SMEFT analyses, complementary to the experimental inputs.

High energy lepton colliders can probe these operators even further.  
The precision reach on the parameter $a$ scales with the energy $E$ and luminosity $L$ as $\Delta a \sim E^{-3} L^{-1/2} $, as the energy dependence of the dim-8 contribution gives an $E^{-4}$ dependence, and the measurement uncertainties are proportional to $(\sigma_{\rm SM} \cdot L)^{-1/2} \sim E \cdot L^{-1/2}$.  
%
Since $a \sim 1/\Lambda^4$, the reach of $\Lambda$ thus scales as
\begin{equation}
\frac{\Lambda_2}{\Lambda_1} = \left( \frac{E_2}{E_1} \right)^{\frac{3}{4}} \left( \frac{L_2}{L_1} \right)^{\frac{1}{8}} \,,  \label{eq:Lambdascale}
\end{equation}
assuming all other variables 
are the same for the two scenarios 1 and 2.  

\begin{figure}
\centering
\includegraphics[width=0.48\textwidth]{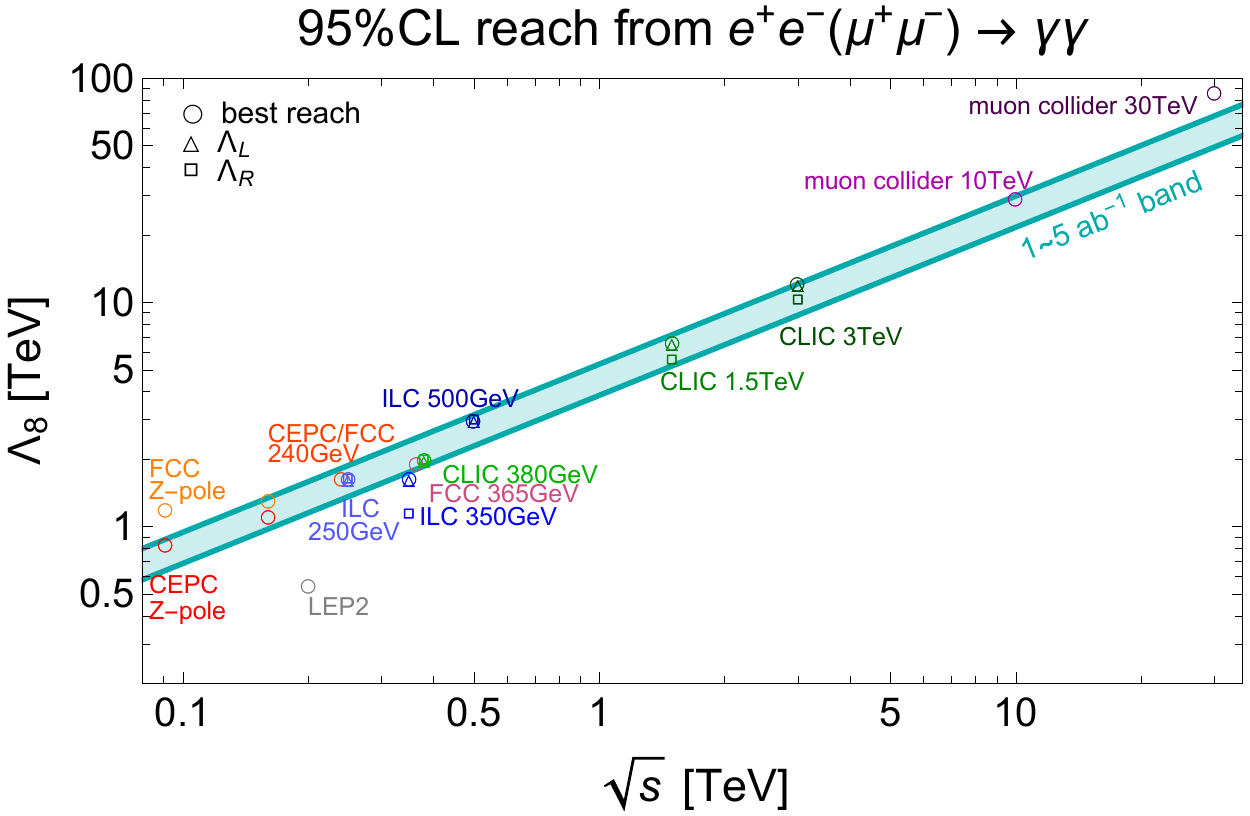}
\caption{The reach on the scale of the dim-8 operators $\Lambda_8 (\equiv
	v/a^{\frac{1}{4}})$ as a function of the center-of-mass energy
	$\sqrt{s}$ from the measurement of the $\eeaa$ (or $\mumuaa$) process.
	The band covers $1-5\inab$ and various beam polarization scenarios. 
	The circle represents the best reach of each collider scenario. 
The LEP2~\cite{Schael:2013ita} reach is shown assuming a SM central value.  
For linear colliders, the triangle (square) shows the
reach for $a_L$ ($a_R$) in a simultaneous fit of both parameters. 
Note that the luminosities of FCC $Z$-pole (150$\inab$), LEP2 (3$\infb$) and muon collider 30\,TeV (90$\inab$) are all very different from the 1-5$\inab$ of the band.
}
\label{fig:LambdaEaa1}
\end{figure}

In \autoref{fig:LambdaEaa1}, we show the $95\%$ CL reach for $\Lambda_8
\equiv v/a^{\frac{1}{4}}$ for various collider scenarios, where $a=a_L,\,a_R$
is defined in Eqs.~(\ref{eq:Aeeaa}).  
$\Lambda_8$ corresponds directly to the scale of new physics which modifies the
$\eeaa$ amplitudes.  
The band 
covers integrated luminosities of $1$ to $5\inab$ and various beam polarization
scenarios, and is consistent with Eq.~(\ref{eq:Lambdascale}). 
We also show the best reach for each collider scenario listed in Appendix~\ref{sec:app} 
from any linear combinations of $a_L$ and $a_R$.  
%
For linear colliders, $a_L$ and $a_R$ can be independently constrained, and the corresponding $\Lambda_L$ and $\Lambda_R$ are also shown.
%

Similar analyses can be carried out for muon colliders,
which probe operators associated with muon fields. 
Constraints on muon dipole moments~\cite{Bennett:2008dy} are
significantly weaker than those of the electron ones. 
We find that two insertions of the electric dipole operator could generate a deviation 
in the $\mumuaa$ cross section comparable to the expected precision reach.  
However, future improvements on the muon electric dipole moment measurement
could make this contribution irrelevant 
($\frac{\delta\sigma_{\gamma\gamma}}{\sigma_{\gamma \gamma }}  \sim (\frac{E}{10^5 {\rm TeV}})^2$)
\cite{Crivellin:2018qmi, Abe:2019thb}.  On the contrary, the current
muon $g_{\mu}-2$ measurement~\cite{Bennett:2006fi,Zyla:2020zbs,Muong-2:2021ojo} sufficiently
constrains the magnetic dipole operator, so that the latter can be safely
ignored for the diphoton measurement
($\frac{\delta\sigma_{\gamma\gamma}}{\sigma_{\gamma \gamma }}  \sim (\frac{E}{10^5 {\rm TeV}})^2$),
independent of whether the apparent discrepancy with the SM is
confirmed.

\sec{Interplay with $Z\gamma$ and $ZZ$ measurements} 
The same operators that enter Eq.~(\ref{eq:aLRop}) also contribute to the
$Z\gamma$ and $ZZ$ processes. 
These processes are, however, more complicated due to
the massive $Z$ boson, which enables contributions to multiple helicity states
from both SM and dim-8 operators (including those responsible for neutral
triple-gauge-boson
couplings~\cite{Gounaris:1999kf,Degrande:2013kka,Ellis:2019zex,Ellis:2020ljj}).
Dim-6 operators could also contribute at the tree level via modifications of
the $Ze^+e^-$ couplings.  At very high energies ($\sqrt{s}\gg m_Z$),
the $Z$ boson is effectively massless, and the $+-$ final helicity states dominate
the $Z\gamma$ and $ZZ$ cross sections \cite{Bellazzini:2018paj}. In this limit,
the $ZZ$ process also exhibits a similar positivity bound, 
\begin{equation}
\frac{d \sigma}{d|\!\cos\theta|} (e^+e^- \to ZZ) \ge \frac{d  \sigma_{\rm SM}}{d|\!\cos\theta|} (e^+e^- \to ZZ) \,. \label{eq:posxzz}
\end{equation}
For the $Z\gamma$ process, we focus on the CP-even elastic amplitude $\A(eV\to eV)$, where $V$ is an arbitrarily mixed state
of $\gamma$ and $Z$.  This gives
%
%
\begin{equation}
(a^{Z\gamma}_L)^2 \le a^{ZZ}_L a^{\gamma\gamma}_L \,, ~~~~ (a^{Z\gamma}_R)^2 \le a^{ZZ}_R a^{\gamma\gamma}_R \,, \label{eq:zgammabound}
\end{equation}
where $a^{Z\gamma}_{L,R}$ ($a^{ZZ}_{L,R}$) is defined as in
Eq.~(\ref{eq:Aeeaa}) with $\gamma^+\gamma^-$ replaced by $Z^+\gamma^-$
($Z^+Z^-$), together with $a_{L,R} \to a^{\gamma\gamma}_{L,R}$ to distinguish
them.  
This implies a simple relation among the $Z\gamma$, $\gamma\gamma$ and $ZZ$ cross sections for any fixed collider scenario (again only in the $\sqrt{s}\gg m_Z$ limit), 
\begin{equation}
(\Delta \sigma_{Z\gamma})^2 \le 4 \Delta \sigma_{\gamma\gamma} \Delta \sigma_{ZZ} \,,  \label{eq:posxza}
\end{equation}
where $\Delta \sigma \equiv \frac{d \sigma}{d|\!\cos\theta|} -\frac{d  \sigma_{\rm SM}}{d|\!\cos\theta|}$. 
We note here again that a proper treatment of the $Z\gamma$ and $ZZ$ processes requires the inclusion of all helicity states of the gauge bosons.  The decay of the $Z$ boson also provides new observables sensitive to the interference of different $Z$ helicity states~\cite{Azatov:2017kzw, Panico:2017frx, Ellis:2020ljj}.  
The mapping between positivity bounds and observables in the $Z\gamma/ZZ$ processes are generally more complicated, and we leave such an analysis to future studies.   On the contrary, the positivity bound of the diphoton process is simple and unambiguous, as we emphasized above.  

\sec{Violation of positivity} 
The observation of 
$\frac{d \sigma  (\eeaa)}{d|\!\cos\theta|} <  \frac{d  \sigma_{\rm SM}  (\eeaa) }{d|\!\cos\theta|}$
does not
necessarily establish the violation of positivity bounds. It is important to
check whether the EFT description itself is invalid, for
instance, due to contributions from new light particles.    In
this process, however, it is difficult for such particles to generate a sizable
destructive interference term while evading the current and future search
constraints. A $t$-channel fermion exchange only generates a
constructive interference. Another possibility is an $s$-channel exchange of a
light composite spin-2 particle, which could be very-well probed by the
resonance search $e^+e^- \to X \gamma / XZ, ~ X\to
\gamma\gamma/e^+e^-$~\cite{Fujii:2016raq}.  Measuring the diphoton process at
multiple center-of-mass energies also helps probe or exclude these light
particle contributions 
and further verifies that the observed deviations are generated by dim-8 operators.  
After all other possibilities are excluded, the result
would then indicate the breakdown of the fundamental principles of
QFT~\footnote{``When you have excluded the impossible, whatever remains,
however improbable, must be the truth.'' -- Sherlock Holmes}.
Interestingly, a recent study~\cite{Aoki:2021ffc} shows with an explicit example that  order-one violations of positivity bounds could be generated if the Lorentz symmetry is spontaneously broken. 

\sec{Summary and outlook} Positivity bounds
require that the diphoton cross section at lepton colliders must be no smaller than
the SM prediction, which offers a rare opportunity to clearly and unambiguously
test the fundamental principles of QFT. 
While high energy colliders provide
the best reaches, such probes are 
robust even for a collider at 
around 240-250\,GeV, a feature that is unique
for the diphoton process.
Alternatively, imposing these bounds could lift the flat directions 
among operators, indicating that 
positivity could provide important information for future global analyses with
dim-8 operators. 

Hadron colliders, such as the LHC or a future 100-TeV collider, have a large
center-of-mass energy and could potentially provide 
powerful probes on the dim-8 operators and their associated positivity bounds~\cite{Bellazzini:2018paj,Bi:2019phv,Alioli:2020kez}.
In particular, a similar process with quarks, $q\bar{q}\to \gamma\gamma$ (or $q\bar{q}\to Z\gamma/ZZ$ \cite{Bellazzini:2018paj}), is already probed at the LHC with a larger center-of-mass energy than the ones of most future lepton colliders. 
However, 
these measurements usually suffer from low measurement precisions which make the EFT interpretation problematic, 
and a consistent EFT treatment often results in much reduced sensitivities to the new physics scale~\cite{Contino:2016jqw, Alte:2017pme}. 
This is particularly important for probing the positivity bounds, for which the contributions of dim-10 operators, not subject to the same bounds, are a potential source of contamination.  
On the other hand, a potential future high energy photon
collider~\cite{ALEGRO:2019alc} could measure the reverse process $\gamma\gamma
\to f \bar{f} $ for different fermion final states, and probe a wider range of
operators and their associated positivity bounds.  We leave the detailed
analyses of these colliders to future studies.

\bigskip

{\noindent \bf Acknowledgement}~~We thank Ying-Ying Li for useful discussions.  JG is supported by the Cluster of Excellence ``Precision Physics, Fundamental Interactions, and Structure of Matter'' (PRISMA+ EXC 2118/1) funded by the German Research Foundation (DFG) within the German Excellence Strategy (Project ID 39083149).  LTW is supported by the DOE grant DE-SC0013642. 
CZ is supported by IHEP under Contract No.~Y7515540U1, and by National Natural Science Foundation of China (NSFC) under grant No.~12075256 and No.~12035008.


\setcounter{secnumdepth}{2}

\appendix

\section{}
\label{sec:app}

\begin{figure}[t]
\centering
\includegraphics[width=0.2\textwidth]{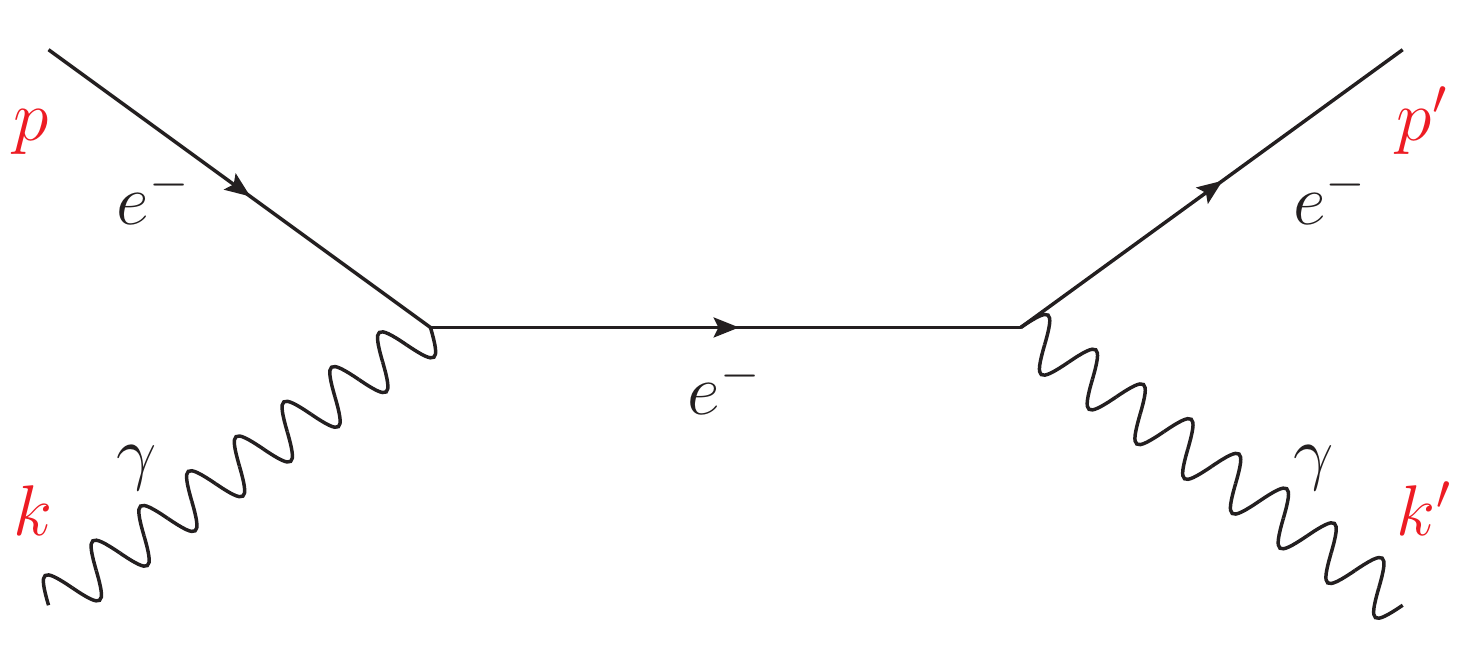} \hspace{0.2cm}
\includegraphics[width=0.2\textwidth]{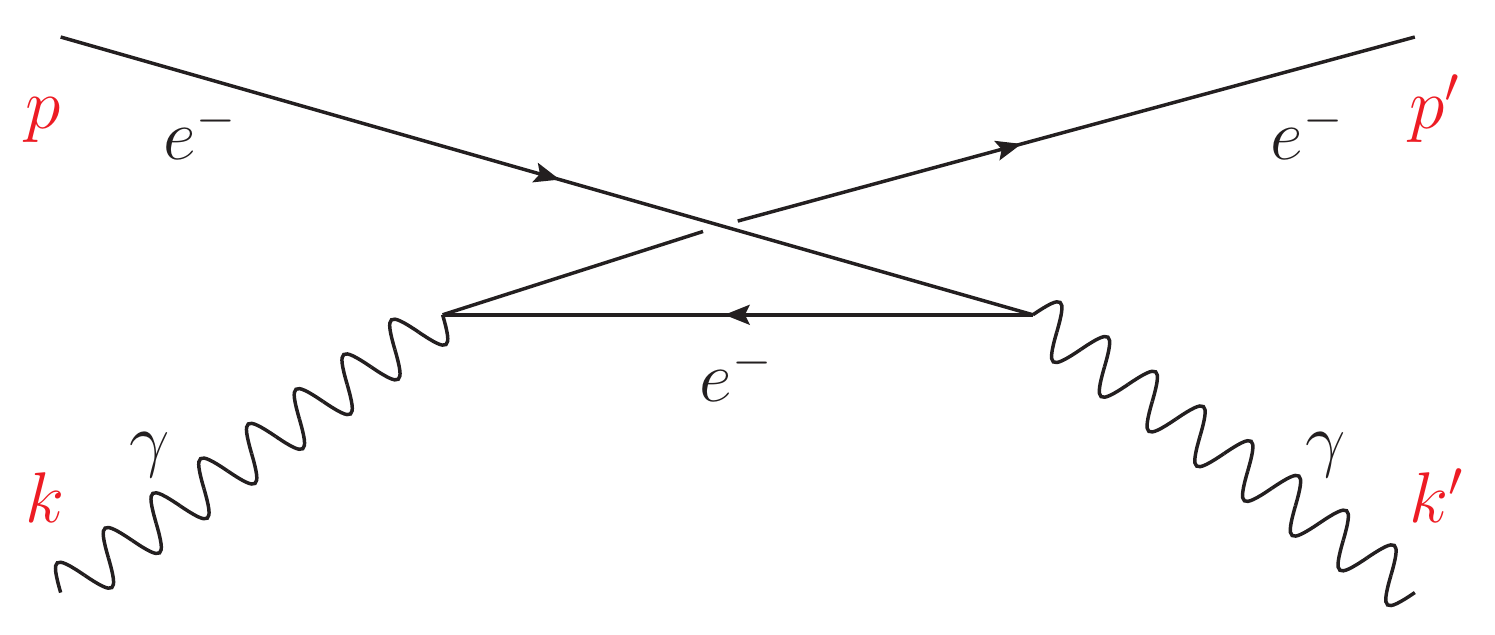}
\caption{Feynman diagrams for $e^- \gamma \to e^- \gamma$ in the SM.}
\label{fig:ea1}
\end{figure}

\begin{figure}[t]
\centering
\includegraphics[width=0.47\textwidth]{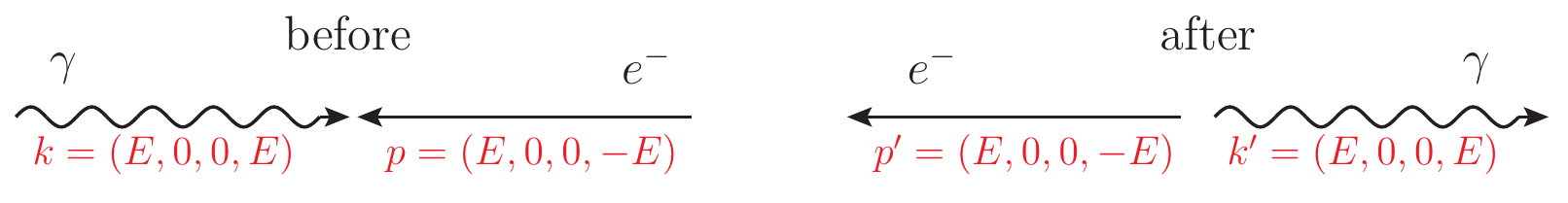}
\caption{Momenta of the forward scattering process before and after the collision.}
\label{fig:ea3}
\end{figure}

\noindent {\bf SM $e^- \gamma \to e^- \gamma$ in the forward limit:} 
We perform an explicit calculation of the amplitude $\M (e^- \gamma \to e^- \gamma)$ in SM in the forward limit with massless electrons.  Our calculation follow closely Sec~5.5 of Ref.~\cite{Peskin:1995ev}.  The general amplitude is given by 
\begin{equation}
\small
i \M = -i e^2 \epsilon^*_\mu (k') \epsilon_\nu (k) \bar{u}(p') \!  \left[ \frac{\gamma^\mu (\slashed{p}+\slashed{k})\gamma^\nu}{(p+k)^2} + \frac{ \gamma^\nu (\slashed{p}-\slashed{k}') \gamma^\mu}{(p-k')^2}  \right] \! u(p) \,, \label{eq:eaM1}
\end{equation}
where $p$, $k$, $p'$ and $k'$ are the momenta of the ingoing $e^-$, $\gamma$ and the outgoing $e^-$, $\gamma$, respectively, as shown in \autoref{fig:ea1}.  
We assume that both $e^-$ and $\gamma$ have $+$ helicity (right-handed).  It can be shown that the results are the same for the other helicity configurations.  To calculate the amplitude in the forward limit, it is most convenient to choose a particular reference frame and a basis for the spinors.  We choose the initial (and final) momenta to be along the $z$-axis, as shown in \autoref{fig:ea3}.  
The Dirac matrices are given by
\begin{equation}
\gamma^\mu = \bpm 0 & \sigma^\mu \\  \bar{\sigma}^\mu & 0 \epm \,,  \hspace{0.2cm} \mbox{ where }  \hspace{0.2cm} \sigma^\mu = (1, \vec{\sigma})\,,~~~  \bar{\sigma}^\mu = (1, -\vec{\sigma}) \,.
\end{equation}
Since we have chosen $e^-$ and $\gamma$ to be right-handed, we have for the Dirac spinor
\begin{equation}
u(p) = \bpm 0 \\ u_R (p) \epm  \hspace{0.2cm} \mbox{ where }  \hspace{0.2cm} u_R (p) = u_R (p') = \sqrt{2E} \bpm 0 \\ 1 \epm \,,
\end{equation}
%
following the spinor choice in Ref.~\cite{Peskin:1995ev}. 
This also gives
\begin{equation}
\bar{u}(p') \gamma^\mu = u^\dagger(p') \gamma^0 \gamma^\mu= \bpm 0 & u^\dagger_R(p) \sigma^\mu \epm \,.  
\end{equation}
The polarization vector are given by
\begin{equation}
\epsilon^\mu = \frac{1}{\sqrt{2}} (0,~1,~i,~0)^\mu \,, \hspace{0.3cm}  \epsilon^{*\mu} = \frac{1}{\sqrt{2}} (0,~1,~ -i,~0)^\mu \,,
\end{equation}
which gives 
\begin{equation}
\sigma^\mu \epsilon_\mu = \bpm 0 & -\sqrt{2} \\ 0 & 0 \epm  \,, \hspace{0.3cm} 
\sigma^\mu \epsilon^*_\mu = \bpm 0 & 0 \\  -\sqrt{2} & 0 \epm \,.
\end{equation}
Plugging everything into \autoref{eq:eaM1}, we have
\begin{align}
\M =& -e^2 \epsilon^*_\mu  \epsilon_\nu u^\dagger_R \left[  \frac{\sigma^\mu \bar{\sigma}\cdot (p+k)\sigma^\nu}{s}   +  \frac{\sigma^\nu \bar{\sigma}\cdot (p-k')\sigma^\mu}{u} \right] u_R \nonumber\\
=& -e^2 \epsilon^*_\mu  \epsilon_\nu 2E  \bpm 0 & 1\epm  \!\! \left[ \! \bpm 0 & 0 \\ -\sqrt{2} & 0 \epm \!\! \frac{2E}{4E^2}  \!\! \bpm 0 & -\sqrt{2}  \\  0 & 0  \epm + 0 \right] \!\!  \bpm 0 \\ 1\epm \nonumber \\
=&   -2e^2 \,.
\end{align}
%

%

\vspace{0.5cm}


\noindent {\bf Operators:} Eq.~\ref{eq:Q} shows the five dimension-8 operators mentioned in the main text, as defined in Ref.~\cite{Murphy:2020rsh}.  The lepton flavor indices are omitted as they are not relevant for our study.  The Lagrangian is written in the form $\mathcal{L}_{\rm dim\mbox{-}8}= \underset{i}{\sum}{ \frac{c_i}{\Lambda^4} Q_i}$.  
\begin{align}
Q_{l^2 B^2 D} =&~ i(\bar{l}\gamma^\mu \overleftrightarrow{D}^\nu l) B_{\mu\rho} B_\nu^{~\rho} \,, \nonumber \\ 
Q^{(2)}_{l^2 WB D} =&~ i(\bar{l}\gamma^\mu \tau^I \overleftrightarrow{D}^\nu l) (B_{\mu\rho} W_\nu^{I\rho}+ B_{\nu\rho} W_\mu^{I\rho}) \,, \nonumber \\
Q^{(1)}_{l^2 W^2 D} =&~ i(\bar{l}\gamma^\mu \overleftrightarrow{D}^\nu l) W^I_{\mu\rho} W_\nu^{I\rho} \,, \nonumber \\
Q_{e^2 B^2 D} =&~ i(\bar{e}\gamma^\mu \overleftrightarrow{D}^\nu e) B_{\mu\rho} B_\nu^{~\rho} \,, \nonumber \\ 
Q_{e^2 W^2 D} =&~ i(\bar{e}\gamma^\mu \overleftrightarrow{D}^\nu e) W^I_{\mu\rho} W_\nu^{I\rho} \,. \label{eq:Q}
\end{align}
%



\begin{figure}[h!]
\centering
\includegraphics[width=0.31\textwidth]{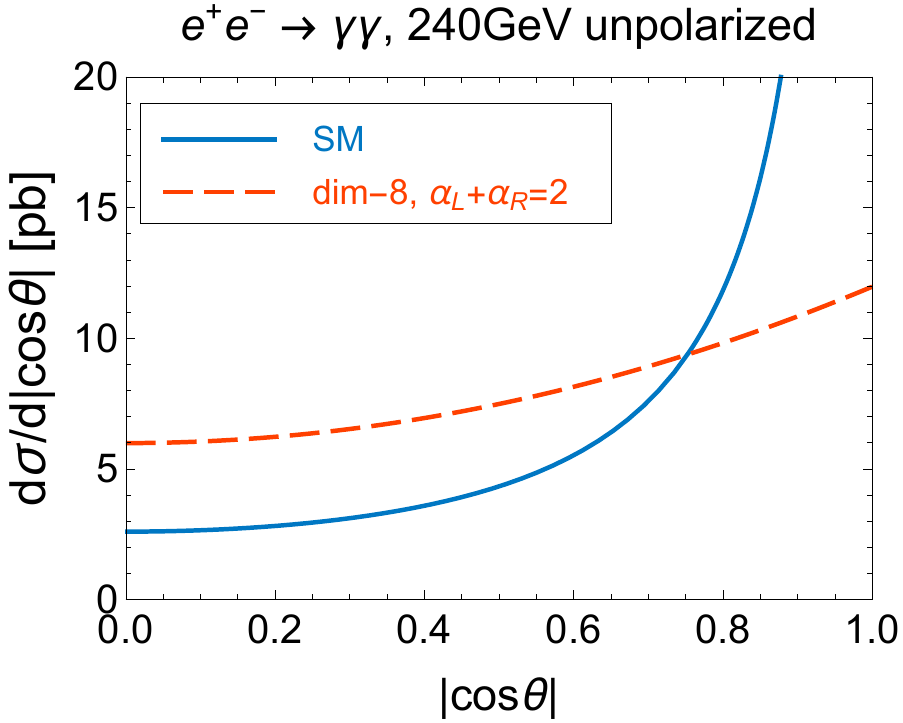}
\caption{The differential cross section $d\sigma/d|\!\cos\theta|$ for the SM and the d8 contribution (as in Eq.~(4) in the main text) for $\sqrt{s}=240\,$GeV, unpolarized beams.  For the d8 contribution, an arbitrary benchmark of $a_L+a_R=2$ is chosen.}
\label{fig:ddcos2}
\end{figure}

\noindent {\bf Differential cross section:} \autoref{fig:ddcos2} shows the differential cross section $d\sigma/d|\!\cos\theta|$ for the SM and the d8 contribution for $\sqrt{s}=240\,$GeV, unpolarized beams. The SM contribution dominates in the forward region due to the $t/u$-channel electron exchange.

%


 \begin{table}[h!]
 \small
 \centering
 \begin{adjustbox}{max width=0.37\textwidth}
 \begin{tabular}{|c||c|c|c|c|} \hline
     \multicolumn{5}{|c|}{ $\int \mathcal{L} dt ~~[\inab]$ }  \\ \hline \hline
unpolarized  & 91\,GeV & 161\,GeV  & 240\,GeV & 365\,GeV    \\ \hline
  CEPC &  8 & 2.6 & 5.6  &       \\ \hline
  FCC-ee & 150 & 10 &  5 &  1.5  \\ \hline\hline
  ILC     &    250\,GeV & 350\,GeV &  500\,GeV &  \\ \hline
  {\scriptsize $(-0.8,+0.3)$}     &   0.9  & 0.135  &  1.6 & \\
   {\scriptsize $(+0.8,-0.3)$}     &   0.9  & 0.045 &  1.6  & \\
   {\scriptsize $(\pm0.8,\pm0.3)$}     &   0.1  & 0.01 &  0.4  & \\\hline     \hline
  CLIC    &       380\,GeV  &  1.5\,TeV &  3\,TeV &  \\ \hline
   {\scriptsize $(-0.8,0)$}   &      0.5  &  2  &  4 & \\
   {\scriptsize $(+0.8,0)$}   &     0.5  &  0.5  &  1 & \\ \hline \hline
 muon collider    & 10\,TeV & 30\,TeV & &  \\ \hline
 unpolarized &  10  &   90  & &   \\ \hline
\end{tabular}
 \end{adjustbox}
\caption{ A summary of the run scenarios of future lepton colliders considered in our analysis with the corresponding integrated luminosity.}
 \label{tab:scenarios}
 \end{table}
 
\vspace{0.5cm} 
 
\noindent {\bf Run scenarios:} \autoref{tab:scenarios} is a summary of the run scenarios of future lepton colliders considered
in our analysis, with the corresponding integrated luminosity.  For ILC and
CLIC, the numbers in the brackets are the values of beam polarizations
$P(e^-,e^+)$.  For simplicity, we assume the $Z$-pole or $WW$-threshold runs
are at one single energy.  Numbers are taken from Refs.~\cite{deBlas:2019wgy,
Han:2020pif}.  Further  details  can be found in
Refs.~\cite{CEPCStudyGroup:2018ghi,Abada:2019lih,Abada:2019zxq,Bambade:2019fyw,deBlas:2018mhx,Delahaye:2019omf}. 

\vspace{0.5cm}

\begin{figure}[t]
\centering
\includegraphics[width=0.37\textwidth]{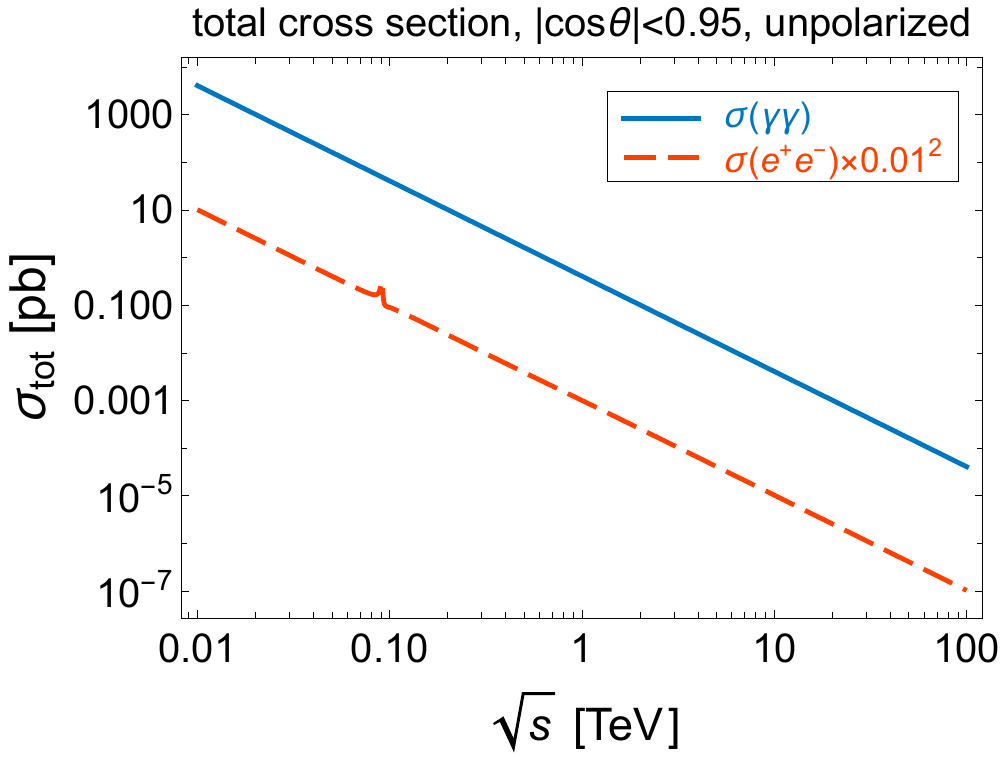}
\caption{The total cross sections of $\eeaa$ and the main background from $e^+e^-\to e^+e^-$ with both final state particles mistagged as a photon.  A cut on the production polar angle $|\!\cos\theta|<0.95$ is applied.  A conservative $1\%$ mistag rate is assumed for both electrons and positrons.}
\label{fig:xsb1}
\end{figure}

\noindent {\bf Measurement uncertainties:}  We have only considered statistical uncertainties in our analysis.  Here we provide further verifications of this assumption.  \autoref{fig:xsb1} shows the total cross sections of diphoton process and the main background from Bhabha scattering ($e^+e^-\to e^+e^-$), the latter contributes to the diphoton channel if both final state particles mistagged as a photon.  Even with a conservative $1\%$ mistag rate for both electrons and positrons, this background is more than two orders of magnitude smaller than the signal.  This is consistent with the LEP analysis in {\it e.g.} Ref.~\cite{L3:1995nbq}, which stated that the contamination from the major background, Bhabha events, was estimated to be less than 0.5\% after selection cuts. 
Another potential source of background is the double-hard-FSR: $e^+e^-\to e^+e^-
\gamma \gamma$, where the two photons take most of the energy of the
scattered electrons. Applying $m_{\gamma\gamma}\ge 0.9s$
on the invariant mass of the two photons with other reasonable cuts,
we estimate the cross section of this
process to be $6\sim7$ orders of magnitude lower than the signal process.  
The statistical uncertainties of the total diphoton cross section measurements are provided for each collider scenario in \autoref{tab:stun} as references.

%

 
 \begin{table}[h!]
 \small
 \centering
 \begin{adjustbox}{max width=0.45\textwidth}
 \begin{tabular}{|c||c|c|c|c|} \hline
     \multicolumn{5}{|c|}{ $\Delta \sigma_{\rm tot}/\sigma_{\rm tot}$, $\eeaa$ }  \\ \hline \hline
unpolarized  & 91\,GeV & 161\,GeV  & 240\,GeV & 365\,GeV    \\ \hline
  CEPC &  $5.1\times 10^{-5}$ & $1.6\times10^{-4}$ & $1.6\times10^{-4}$  &       \\ \hline
  FCC-ee & $1.2\times 10^{-5}$ & $8.0\times 10^{-5}$ & $1.7\times10^{-4}$  & $4.7\times10^{-4}$   \\ \hline\hline
  ILC     &    250\,GeV & 350\,GeV &  500\,GeV &  \\ \hline
      &  $2.5\times10^{-4}$   & $1.1\times10^{-3}$  & $3.7\times10^{-4}$  & \\  \hline\hline
  CLIC    &       380\,GeV  &  1.5\,TeV &  3\,TeV &  \\ \hline
    &    $6.0\times 10^{-4}$    &  $1.5\times 10^{-3}$  & $2.1\times 10^{-3}$  & \\ \hline \hline
 muon collider    & 10\,TeV & 30\,TeV & &  \\ \hline
 &   $5.0\times 10^{-3}$  &   $5.0\times 10^{-3}$  & &   \\ \hline
\end{tabular}
 \end{adjustbox}
\caption{ The projected (relative) statistical uncertainties of the total cross section measurement of $\eeaa$ with the run scenarios in \autoref{tab:scenarios}.  A cut on the production polar angle $|\!\cos\theta|<0.95$ is applied.}
 \label{tab:stun}
 \end{table}

\bibliographystyle{JHEP}
\bibliography{pos2}

\end{document}